\documentclass[runningheads]{llncs}

\usepackage{amsmath,amsfonts,amssymb}
\usepackage{newpxtext} 
\usepackage[normalem]{ulem}
\newcommand*{\knnembed}{\textit{k}NN-Embed}

\DeclareMathOperator*{\argmax}{arg\,max}

\DeclareMathOperator{\centroid}{centroid}

\usepackage{subcaption}
\usepackage{booktabs}

\usepackage{algorithm}
\usepackage{algorithmic}

\usepackage{graphicx}

\usepackage{etoolbox}

\usepackage{kantlipsum} %

\newcommand{\repthanks}[1]{\textsuperscript{\ref{#1}}}
\makeatletter
\patchcmd{\maketitle}
  {\def\thanks}
  {\let\repthanks\repthanksunskip\def\thanks}
  {}{}
\patchcmd{\@maketitle}
  {\def\thanks}
  {\let\repthanks\@gobble\def\thanks}
  {}{}
\newcommand\repthanksunskip[1]{\unskip{}}
\makeatother

\begin{document}
\title{$\mathbf{k}$NN-Embed: Locally Smoothed Embedding Mixtures For Multi-interest Candidate Retrieval}

\titlerunning{kNN-Embed}

\author{Ahmed El-Kishky\thanks{Corresponding author}
 \and
Thomas Markovich \and Kenny Leung \and Frank Portman \and Aria Haghighi\thanks{Equal contribution \protect\label{X}} \and Ying Xiao\repthanks{X}}
\authorrunning{A. El-Kishky et al.}

\institute{Twitter Cortex, San Francisco CA 94103, USA\\
\email{\{aelkishky,tmarkovich,kennyleung,fportman,ahaghighi,yxiao\}@twitter.com}}

\maketitle   
\begin{center}
  \Large\bfseries\boldmath
  $\mathbf{k}$NN-Embed: Locally Smoothed Embedding Mixtures For Multi-interest Candidate Retrieval
\end{center}

\begin{abstract}
Candidate retrieval is the first stage in recommendation systems, where a light-weight system is used to retrieve potentially relevant items for an input user. These candidate items are then ranked and pruned in later stages of recommender systems using a more complex ranking model. As the top of the recommendation funnel, it is important to retrieve a high-recall candidate set to feed into downstream ranking models. A common approach is to leverage approximate nearest neighbor (ANN) search from a single dense query embedding; however, this approach this can yield a low-diversity result set with many near duplicates. As users often have multiple interests, candidate retrieval should ideally return a diverse set of candidates reflective of the user's multiple interests. To this end, we introduce \knnembed, a general approach to improving diversity in dense ANN-based retrieval. \knnembed~represents each user as a smoothed mixture over learned item clusters that represent distinct ``interests'' of the user. By querying each of a user's mixture component in proportion to their mixture weights, we retrieve a high-diversity set of candidates reflecting elements from each of a user's interests. We experimentally compare \knnembed~ to standard ANN candidate retrieval, and show significant improvements in overall recall and improved diversity across three datasets. Accompanying this work, we open source a large Twitter follow-graph dataset\footnote{\label{note2}\url{https://huggingface.co/datasets/Twitter/TwitterFollowGraph}}, to spur further research in graph-mining and representation learning for recommender systems.

\end{abstract}
\section{Introduction}
Recommendation systems for online services such as e-commerce or social networks present users with suggestions in the form of ranked lists of items~\cite{covington2016deep}. Often, these item lists are constructed through a two-step process: (1) candidate retrieval, which efficiently retrieves a manageable subset of potentially relevant items, and (2) ranking, which applies a computationally-expensive ranking model to score and select the top-$k$ candidates to display to the user. 

During candidate retrieval, we are primarily concerned with the \emph{recall} of the system \cite{huang2020embedding}, as opposed to the ranking model which typically targets \emph{precision}. Ensuring high recall for users with multiple interests is a challenging problem, which is exacerbated by the way we typically perform retrieval. The dominant paradigm for candidate retrieval is to embed users and items in the same vector space, and then use approximate nearest-neighbor (ANN) search to retrieve candidates close to the user~\cite{kang2019candidate,covington2016deep}. However, ANN search will often return  candidate pools that are highly intra-similar (e.g., all candidates pertain to one ``topic'' only)~\cite{wilhelm2018practical}. A side effect of training embeddings to place users close to relevant items, is that similar items are also placed close to each other. During ANN-based candidate retrieval, this unfortunately leads to similar candidates that may not reflect a user's diverse multi-topic interests, and hence low recall.

In this paper, we introduce \knnembed, a new strategy for retrieving a high-recall, diverse set of candidates reflecting a user's multiple interests. \knnembed~ captures multiple user interests by representing user preferences with a smoothed, mixture distribution. Our technique provides a turn-key way to increase recall and diversity while maintaining user relevance in any ANN-based candidate retrieval scheme. It does not require retraining the underlying user and item embeddings; instead, we build directly on top of pre-existing ANN systems. The underlying idea is to exploit the similarity of neighboring users to represent per-user interests as a mixture over learned high level clusters of item embeddings. Since user-item relevance signal is typically sparse, estimating the mixture weights introduces significance variance. Thus, we smooth the mixture weights with information from similar users. At retrieval time, we simply sample candidates from each cluster according to mixture weights. Within each cluster, we perform ANN search using a smoothed per-user per-cluster embedding. 

Our contributions in this paper are (1) a principled method to retrieve a high-recall, diverse candidate set in ANN-based candidate retrieval systems and (2) a large open-source graph dataset for studying graph-mining and retrieval.

\section{Related Works}\label{sec:related}
Traditionally, techniques for candidate retrieval rely on fast, scalable approaches to search large collections for similar sparse vectors~\cite{bayardo2007scaling,andoni2006near}. Approaches apply indexing and optimization strategies to scale sparse similarity search. One such strategy builds a static clustering of the entire collection of items; clusters are retrieved based on how well their centroids match the query~\cite{van1975document,liu2004cluster}. These methods either (1) match the query against clusters of items and rank clusters based on similarity to query or (2) utilize clusters as a form of item smoothing. %

For embedding-based recommender systems~\cite{zhang2016collaborative}, large-scale dense similarity search has been applied for retrieval. Some approaches proposed utilize hashing-based techniques such as mapping input and targets to discrete partitions and selecting targets from the same partitions as inputs~\cite{weston2013label}. With the advent of fast approximate nearest-neighbor search~\cite{malkov2018efficient,johnson2019billion}, dense nearest neighbor has been applied by recommender systems for candidate retrieval~\cite{covington2016deep}.

When utilizing graph-based embeddings for recommender systems~\cite{el2022twhin}, some methods transform single-mode embeddings to multiple modes by clustering user actions~\cite{pal2020pinnersage}. Our method extends upon this idea by incorporating nearest neighbor smoothing to address the sparsity problem of generating mixtures of embeddings for users with few engagements.

Smoothing via k-nearest-neighbor search has been applied for better language modeling~\cite{khandelwal2019generalization} and machine translation~\cite{khandelwal2020nearest}. We smooth low-engagement user representations by leveraging engagements from similar users.

\section{\knnembed}
\subsection{Preliminaries}
\label{sec:preliminaries}
Let $\mathcal{U} = \{u_1, u_2, \ldots u_n\}$ be the set of source entities (i.e., users in a recommender system) and $\mathcal{I} = \{i_1, i_2, \ldots i_m\}$ be the set of target entities (i.e., items in a recommender system). Let $\mathcal{G}$ constitute a bipartite graph representing the engagements between users ($\mathcal{U}$) and items ($\mathcal{I}$). For each user and item, we define a ``relevance'' variable in $\{0, 1\}$ indicating an item's relevance to a particular user. An item is considered relevant to a particular user if a user, presented with an item, will engage with said item. Based on the engagements in $\mathcal{G}$, each user, $u_j$, is associated with  a $d$-dimensional embedding vector $\mathbf{u_j} \in \mathbb{R}^d$; similarly each target item $i_k$ is associated with an embedding vector $\mathbf{i_k} \in \mathbb{R}^d$. We call these  the \emph{unimodal} embeddings, and assume that they model user-item relevance $p(\textrm{relevance} | u_j, i_k) = f(\mathbf{u_j}, \mathbf{i_k})$ for a suitable function $f$.

Given the input user-item engagement graph, our goal is to learn mixtures of embeddings representations of users that better capture the multiple interests of a user as evidenced by higher recall in a candidate retrieval task.

\paragraph{\textbf{Unimodal User and Item Embeddings:}}
While \knnembed~presupposes a set of co-embedded user and item embeddings and is agnostic to the exact embedding technique used (the only constraint is that the embeddings must satisfy $p(i_k|u_j) = g(\mathbf{u_j}^T\mathbf{i_k})$ for monotone $g$), for completeness we describe a simple approach we applied to co-embed users and items into the same space. We form a bipartite graph $\mathcal{G}$ of users and items, where an edge represents relevance (e.g., user follows content producer).  We seek to learn an embedding vector (i.e., vector of learnable parameters) for each user ($u_j$) and item ($i_k$) in this bipartite graph; we denote these learnable embeddings for users and items as $\mathbf{u_j}$ and $\mathbf{i_k}$ respectively. A user-item pair is scored with a scoring function of the form $f(\mathbf{u_j}, \mathbf{i_k})$. Our training objective seeks to learn $\mathbf{u}$ and $\mathbf{i}$ parameters that maximize a log-likelihood constructed from the scoring function for $(u, i) \in \mathcal{G}$ and minimize for $(u, i) \notin \mathcal{G}$. For simplicity, we apply a dot product comparison between user and item representations. For a user-item pair $e=(u_j, i)$, this is defined by:
\begin{equation}
    \label{eq:scoring}
    f(e) = f(u_j, i_k) = \mathbf{u_j}^\intercal \mathbf{i_k}
\end{equation}

As seen in Equation~\ref{eq:scoring}, we co-embed users and items by scoring their respective embedded representations via dot product and perform edge (or link) prediction. We consume the input bipartite graph $\mathcal{G}$ as a set of user-item pairs of the form  $(u, i)$ which represent positive engagements between a user and item. The embedding training objective is to find user and item representations that are useful for predicting which users and items are linked via an engagement. While a softmax is a natural formulation to predict a user-item engagement, it is impractical due to the cost of computing the normalization over a large vocabulary of items. Following previous methods~\cite{mikolov2013distributed,goldberg2014word2vec}, negative sampling, a simplification of noise-contrastive estimation, can be used to learn the  parameters  $\mathbf{u}$ and $\mathbf{i}$. We maximize the following negative sampling objective: 
\begin{equation}
    \label{eq:objective}
    \argmax_{\mathbf{u}, \mathbf{i}}\sum_{e \in \mathcal{G}} \left[  \log \sigma (f(e)) + \sum_{e' \in N(e)} \log \sigma (-f(e')) \right]
\end{equation}
where: $N(u, i) = \{(u, i'): i'\in \mathcal{I}\} \cup \{(u', i): u' \in \mathcal{U}\}$.
Equation~\ref{eq:objective} represents the log-likelihood of predicting a binary ``real" (edges in the network) or ``fake'' (negatively sampled edges) label. To maximize the objective, we learn $\mathbf{u}$ and $\mathbf{i}$  parameters to differentiate positive edges from negative, unobserved edges. Negative edges are sampled by corrupting positive edges via replacing either the user or item in an edge pair with a negatively sampled user or item. Following previous approaches, negative sampling is performed both uniformly and proportional to node prevalence in the training graph~\cite{bordes2013translating,lerer2019pytorch}.

\subsection{Smoothed Mixture of Embeddings}\label{sec:knn_embed}

To use embeddings for candidate retrieval, we need a method of selecting relevant items given the input user. Ideally, we would like to construct a full distribution over all items for each user $p(i_k |u_j)$ and draw samples from it. The sheer number of items makes this difficult to do efficiently, especially when candidate retrieval strategies are meant to be light-weight. In practice, the most common method is to greedily select the top few most relevant items using an ANN search with the unimodal user embedding as query. A significant weakness of this greedy selection is that, by its nature, ANN search will return items that are similar not only to the user embedding, but also to each other; this drastically reduces the \emph{diversity} of the returned items. This reduction in diversity is a side-effect of the way embeddings are trained -- typically, the goal of training embeddings is to put users and relevant items close in Euclidean space; however, this also places similar users close in space, as well as similar items. We will repeatedly exploit this ``locality implies similarity'' property of embeddings in this paper to resolve this diversity issue.

\paragraph{\textbf{Clustering Items:}}
Since neighboring items are similar in the embedding space, if we apply a distance-based clustering to items, we can arrive at groupings that represent individual user preferences well. As such, we first cluster items using spherical k-means~\cite{dhillon2001concept} where cluster centroids are placed on a high-dimensional sphere with radius one. Given these item clusters, instead of immediately collapsing the distribution $p(i_k | u_j)$ to a few items as ANN search does, we can write the full distribution $p(i_k | u_j)$ as a mixture over item clusters:
\begin{align*}
    p(i_k | u_j) = \sum_c p(c | u_j) \cdot p(i_k | u_j, c)
\end{align*}
where in each cluster, we learn a separate distribution over the items in the cluster $p(i_k | u_j, c)$. Thus, we are modeling each user's higher level interests $p(c | u)$, and then within each interest $c$, we can apply an efficient ANN-search strategy as before. In effect, we are interpolating between sampling the full preference distribution $p(i_k | u_j)$ and greedily selecting a few items in an ANN.

\paragraph{\textbf{Mixture of Embeddings via Cluster Engagements:}} After clustering target entities, we learn $p(c | u_j)$ through its maximum likelihood estimator (MLE):
\begin{equation}
    p_{\mathrm{mle}}(c|u_j) = {\mathrm{count}(u_i, c)} / {\sum\limits_{c'\in \mathcal{M}_j} \mathrm{count}(u_j, c')}
    \label{eq:distribution}
\end{equation}
where, $\mathrm{count}(u_j, c)$ is the number of times $u_j$ has a relevant item in cluster $c$. For computational efficiency, we take $\mathcal{M}_j$ to be $u_j$'s top $m$ most relevant clusters. We normalize these counts to obtain a proper cluster-relevance distribution. 

\paragraph{\textbf{Nearest Neighbor Smoothing:}}
Unfortunately, we typically have few user-item engagements on a per-user basis; thus, while the MLE is unbiased and asymptotically efficient, it can also be high variance. To this end, we introduce a smoothing technique that once again exploits locality in the ANN search, this time for users.

\begin{figure}
    \centering
    \includegraphics[scale=0.75]{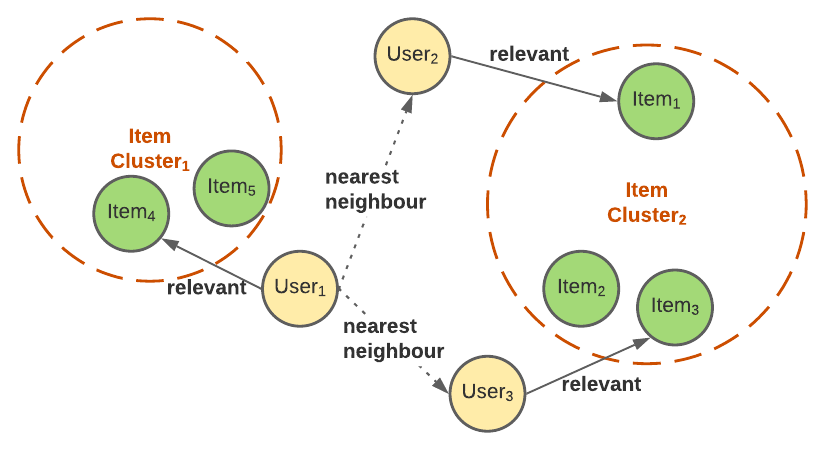}
    \caption{Example of retrieving two candidates. In an ANN, items 4 and 5 would be deterministically returned for user 1. In our proposed \knnembed, even though the distances to cluster 2 are larger, smoothing means that we will sometimes return items from that cluster, yielding more diverse items. Note in this case, we don't even require that user 1 has previously relevant items in cluster 2.}
    \label{fig:diagram}
    \vspace{-0.8cm}
\end{figure}

Figure \ref{fig:diagram} illustrates identifying $k$ nearest-neighbors ($\mathcal{K}_j$) to the query user $u_j$'s, and leveraging the information from the neighbors' cluster engagements to augment the user's cluster relevance. We compute this distribution over item clusters by averaging the MLE probability for each nearest neighbor (item clusters that are not engaged with by a retrieved neighbor have zero probability).

\begin{equation}
    p_{\mathit{k}NN}(c|u_j) = \frac{1}{|\mathcal{K}_j|} \sum_{u' \in \mathcal{K}_j} p_{\mathrm{mle}}(c|u')
    \label{eq:smoothed_distribution}
\end{equation}

We apply Jelinik-Mercer smoothing to interpolate between a user's MLE distribution with the aggregated nearest neighbor distribution~\cite{jelinek1980interpolated}.
\begin{equation}\label{eq:smoothed_probability}
    p_{\mathrm{smoothed}}(c|u_j) =(1-\lambda) p_{\mathrm{mle}}(c|u_j) + \lambda p_{\mathit{k}NN}(c|u_j),
\end{equation}
where $\lambda \in [0, 1]$ represents how much smoothing is applied. It can be manually set or tuned on a downstream extrinsic task.

\ \\
\noindent
\textbf{Sampling within Clusters} Within each cluster there are many ways to retrieve items on a per user basis. A simple, but appealing, strategy is to represent each user as a normalized centroid of their relevant items in that cluster:
\begin{equation}
\label{eq:unsmoothed_cluster_embedding}
    \centroid(c, u_j) =  \frac{\sum_{m \in R(c, u_j)} \mathbf{i_m}}{\|\sum_{m \in R(c, u_j)} \mathbf{i_m} \|} ,
\end{equation}
where $R(c, u_j)$ is the set of relevant items for user $u_j$ in cluster $c$. However, since we are applying smoothing to the cluster probabilities $p(c|u_j)$, it may be case that $u_j$ has zero relevant items in a given cluster. Hence, we smooth the user centroid using neighbor infomation to obtain the final user representation $\mathbf{u_j^c}$:

\begin{equation}
\begin{aligned}[b]
\label{eq:smoothed_cluster_embedding}
     \mathbf{u_j^c} & = (1-\lambda) \centroid(c, u_j)   + \frac{\lambda}{|\mathcal{K}_j|} \sum_{u' \in \mathcal{K}_j} p_{\mathrm{mle}}(c|u') \centroid(c, u')
\end{aligned}
\end{equation}
Equation~\ref{eq:smoothed_cluster_embedding} shows the $\mathbf{k}$NN-smoothed user-specific embedding for cluster $c$. This embedding takes the user-specific cluster representations from Equation~\ref{eq:unsmoothed_cluster_embedding}, and performs a weighted averaging proportionate to each user's contribution to $p_{\mathrm{smoothed}}(c|u_j)$. The final vector is once again normalized to unit norm.

\section{Evaluation Datasets and Metrics}\label{sec:dataset}

We evaluate on three datasets which we describe below:

\paragraph{\textbf{HEP-TH Citation Graph:}} This paper citation network is collected from Arxiv preprints from the High Energy Physics category~\cite{gehrke2003overview}. The dataset consists of: 34,546 papers and 421,578 citations.%

\paragraph{\textbf{DBLP Citation Graph:}}This paper citation network is collected from DBLP~\cite{tang2008arnetminer} and consists of 5,354,309 papers and 48,227,950 citation relationships.%

\paragraph{\textbf{Twitter Follow Graph:}}
We curate Twitter user-follows-user (available via API) by first selecting a number of `highly-followed' users that we refer to as `content producers'; these content producers serve as `items' in our recommender systems terminology. We then sampled users that follow these content producer accounts. All users are anonymized with no other personally identifiable information (e.g., demographic features) present. Additionally, the timestamp of each follow edge was mapped to an integer that respects date ordering, but does not provide any information about the date that follow occurred. In total, we have $261M$ edges and $15.5M$ vertices, with a max-degree of $900K$ and a min-degree of $5$. We hope that this dataset will be of useful to the community as a test-bed for large-scale retrieval research.

\paragraph{\textbf{Metrics:}}
We evaluate \knnembed~ on three aspects: (1) the recall  (2) diversity and (3) goodness of fit of retrieved candidates. Below, we formalize these metrics.

\paragraph{\textbf{Recall@K:}} The most natural (and perhaps most important) metric for computing the efficacy of various candidate retrieval strategies is $Recall@K$. This metric is given by considering a fixed number of top candidates yield by a retrieval system (up to size $K$) and measuring what percent of these candidates are held-out relevant candidates. The purpose of most candidate retrieval systems is to collect a high-recall pool of items for further ranking, and thus recall is a relevant metric to consider. Additionally, recall provides an indirect way to measure diversity -- to achieve high recall, one is obliged to return a large fraction of \emph{all} relevant documents, which simple greedy ANN searches can struggle with.

\paragraph{\textbf{Diversity:}}
To evaluate the diversity among the retrieved candidates, we measure the spread in the embeddings of the retrieved candidates by calculating the average distance retrieved candidates are from their centroid. The underlying idea is that when `locality implies similarity'; as a corollary, if candidates are \emph{further} in Euclidean distance, then they are likely to be different. As such, for a given set of candidates $\mathcal{C}$, we compute diversity $D$ as follows:

\begin{equation}
    D(\mathcal{C}) = \frac{1}{|\mathcal{C}|} \sum_{i_k \in \mathcal{C}}\|\mathbf{i_k}-\hat{\mathbf{i}}\|
    \label{eq:diversity}
\end{equation}
where $\mathcal{C}$ denotes the set of retrieved candidates and $\hat{\mathbf{i}} = \sum_{i_k \in \mathcal{C}} \mathbf{i_k} / |\mathcal{C}|$ is the mean of the unimodal embeddings of the retrieved candidates.%

\begin{figure}
\vspace{-0.7cm}
\centering

  \includegraphics[width=.87\textwidth]{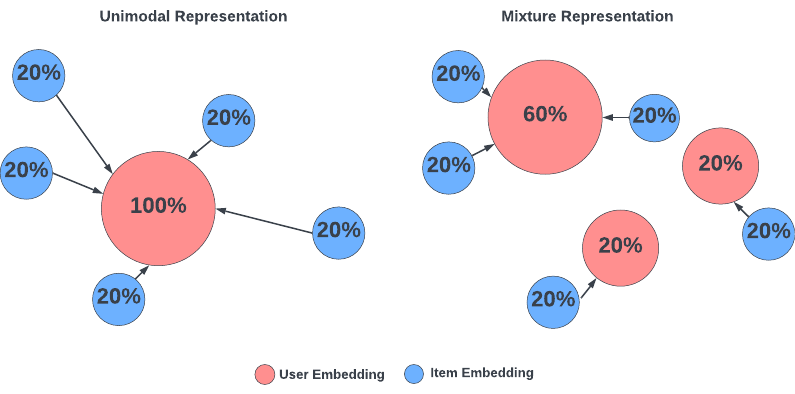}
  \caption{Goodness of fit of unimodal representation vs mixture representation.}
  \label{fig:goodness_fit}
\end{figure}

\paragraph{\textbf{Goodness of Fit:}}
In addition to diversity of retrieved items, we need to ensure that a user's mixture representation is an accurate model of their interests -- that is the mixture of embeddings identifies points in the embedding space where relevant items lie. Thus, we compare held out relevant items to the user's mixture representation we use to query. We measure this ``goodness of fit''  by computing the Earth Mover's Distance (EMD)~\cite{levina2001earth} between a uniform distribution over a user's relevant items and the user's mixture distribution.
The EMD measures the distance between two probability distributions over a metric space~\cite{kusner2015word,el2020massively}. We measure the distance between a user's cluster distribution (e.g.,  Equation~\ref{eq:distribution} and Equation~\ref{eq:smoothed_distribution}), to a uniform distribution over a held-out set of relevant items: $p(i|u_j)$ over a Euclidean space. We compute EMD by soft assigning all held-out relevant items of a user to clusters, minimizing the sum of item-cluster distances, with the constraint that the sum over soft assignments matches $p(c|u_j)$. As seen in Figure~\ref{fig:goodness_fit}, with standard unimodal representations, a single embedding vector is compared to the held-out items and the goodness of fit is the distance between the item embeddings and the singular user embedding. In comparison, for mixture representations (Figure~\ref{fig:goodness_fit}, each user multiple user embeddings who each have fractional probability mass that in total sums to 1. The goodness of fit is then the distance achieved by allocating the mass in each item to the closest user embedding cluster with available probability mass. Observing unimodal representations in Fig.~\ref{fig:goodness_fit}, a single unimodal embedding is situated in the embedding space and compared to held-out relevant items. As shown, some held-out items are close to the unimodal embedding, while others are further away. In contrast, for mixture representations, each user has multiple user-embeddings and each of these embeddings lies close to a cluster of relevant items. The intuition is that if a user has multiple item clusters they are interested in, multiple user embeddings can better capture these interests.

\vspace{-0.9cm}

\section{Experiments}\label{sec:exp}

\paragraph{\textbf{Experimental Setup:}} For our underlying ANN-based candidate retrieval system,  we start by creating a bipartite graph between source entities and target entities for each dataset, with each edge representing explicit relevance between items (e.g., citing paper \textit{cites} cited paper or user \textit{follows} content producer). We then learn unimodal 100-dimensional embeddings for users and items by training over 20 epochs and cluster them via spherical $k$-means over 20 epochs~\cite{arthur2006k}.

\paragraph{\textbf{Evaluation Task:}} We evaluate three candidate retrieval strategies -- baseline ANN with unimodal embeddings (which is how most ANN-based candidate retrieval systems work), mixture of embeddings with no smoothing~\cite{pal2020pinnersage}, and mixture of embeddings with smoothing (i.e., \knnembed). For each strategy, we compute the $Recall@K$, diversity, and fit in a link prediction task.%

\paragraph{\textbf{Research Hypotheses:}} We explore two research hypotheses (as well as achieve some understanding of the hyperparameters): (1) Unimodal embeddings miss many relevant items due to the similarity of retrieved items. Mixtures yield more diverse and higher recall candidates. (2) Smoothing, by using information from neighboring users, further improves the recall of retrieved items.

\setlength{\tabcolsep}{5pt}
\begin{table}
    \vspace{-0.8cm}
    \caption{Recall of Retrieved Candidates}
    \label{tab:recall_tab}
    \begin{tabular}{l |  c c c | c c c | c c c}
    \toprule
    & \multicolumn{3}{c|}{HEP-TH} & \multicolumn{3}{c|}{DBLP} & \multicolumn{3}{c}{Twitter-Follow}\\
    \midrule
    \textbf{Approach} & \textbf{R@10} & \textbf{R@20} & \textbf{R@50}  & \textbf{R@10} & \textbf{R@20} & \textbf{R@50}  & \textbf{R@10} & \textbf{R@20} & \textbf{R@50}\\
    \midrule
    Unimodal & 20.0\% & 30.0\%  & 45.7\%  & 9.4\% & 13.9\%  & 21.6\%  & 0.58\% & 1.02\%  & 2.06\% \\
    Mixture & 22.7\% & 33.4\%  & 49.3\% & 10.9\% & 16.1\%  & 25.1\%  & 3.70\% & 5.53\%  & 8.79\%  \\
    \knnembed & \textbf{25.8\%}  & \textbf{37.4\%}  & \textbf{52.5\%} & \textbf{12.7\%} & \textbf{18.8\%}  & \textbf{28.3\%}  & \textbf{4.13\%} & \textbf{6.21\%}  & \textbf{9.77\%}  \\
    \bottomrule
    \end{tabular}

    \caption*{Recall of unimodal vs mixture vs \knnembed - higher is better. HEP-TH ($\lambda=0.8$, 2000 clusters, 5 embeddings). DBLP ($\lambda=0.8$, 10000 clusters, 5 embeddings). Twitter-Follow ($\lambda=0.8$, 40000 clusters, 5 embeddings).}
    \vspace{-0.8cm}
\end{table}

\paragraph{\textbf{Recall:}}
In Table~\ref{tab:recall_tab}, we report results when evaluating recall on citation prediction tasks. Results support the first hypothesis that unimodal embeddings may miss relevant items if they don't lie close to the user in the shared embedding space. Mixture of embeddings with no smoothing, yields a $14\%$ relative improvement in $R@10$ for for HEP-TH, and $16\%$ relative improvement for DBLP. Our second hypothesis (2) posits that data sparsity can lead to sub-optimal mixtures of embeddings, and that nearest-neighbor smoothing can mitigate this. Our experiments support this hypothesis, as we see a $25\%$ relative improvement for HEP-TH in $R@10$, and $35\%$ for DBLP and  when using \knnembed. We see similar significant improvements over baselines in $R@20$ and $R@50$. For Twitter-Follow, the improvements in recall are dramatic -- $534\%$ in relative terms going from unimodal embeddings to a mixture of embeddings in $R@10$. We suspect this significant improvement is because Twitter-Follow simultaneously has a much higher average degree than HEP-TH and DBLP and the number of unique nodes is much larger. It is a more difficult task to embed so many items, from many different interest clusters, in close proximity to a user. As such, we see a massive improvement by explicitly querying from each user's interest clusters. Applying smoothing provides an additional $74\%$ in relative terms, and similar behaviours are observed in $R@20$ and $R@50$.

\paragraph{\textbf{Diversity:}} We apply Equation~\ref{eq:diversity} to retrieved candidates and measure the spread of retrieved candidates' embedding vectors. As seen in Table~\ref{tab:diversity_tab}, the candidates from unimodal retrieval are less diverse than candidates retrieved via multiple queries from mixture representations. This verifies our first research hypothesis that unimodal embeddings may retrieve many items that are clustered closely together as a by-product of ANN retrieval (i.e., diversity and recall is low). However, multiple queries from mixtures of embeddings broadens the search spatially; retrieved items are from different clusters, which are more spread out from each other. \knnembed~(i.e., smooth mixture retrieval) results in slightly less diverse candidates than unsmoothed mixture retrieval. We posit that this is due to the high-variance of the maximum likelihood estimator of the $p_{\mathrm{mle}}(c|u_j)$ multinomial (Equation~\ref{eq:distribution}). While this high-variance may yield more diverse candidates, this yields less relevant candidates as seen in Table~\ref{tab:recall_tab} where \knnembed~ consistently yields better recall than unsmoothed mixture retrieval. While high diversity is necessary for high recall, it is insufficient on its own.%

\begin{table*}
    \vspace{-0.8cm}
\caption{Diversity of Retrieved Candidates}
\label{tab:diversity_tab}
\begin{tabular}{l | c c c | c c c | c c c}
\toprule
& \multicolumn{3}{c|}{HEP-TH} & \multicolumn{3}{c|}{DBLP} & \multicolumn{3}{c}{Twitter-Follow}\\
\midrule
\textbf{Approach} & \textbf{D@10} & \textbf{D@20} & \textbf{D@50} & \textbf{D@10} & \textbf{D@20} & \textbf{D@50} & \textbf{D@10} & \textbf{D@20} & \textbf{D@50}\\
\midrule
Unimodal & 0.49 & 0.54  & 0.61 & 0.43 & 0.46  & 0.51 & 0.38 & 0.40  & 0.43   \\
Mixture & \textbf{0.58} & \textbf{0.63}  & \textbf{0.68} & \textbf{0.51} & \textbf{0.56}  & \textbf{0.60} & \textbf{0.56} & \textbf{0.54}  & \textbf{0.58}    \\
\knnembed & 0.54 & 0.60  & 0.66  & 0.46 & 0.52  & 0.57 &  0.47 & 0.52  & 0.55   \\
\bottomrule
\end{tabular}
    \vspace{-0.8cm}
\end{table*}

\paragraph{\textbf{Goodness of Fit:}} We evaluate how well unimodal, mixture, and smoothed mixture embeddings model a user's interests. The main idea is that the better fit a user representation is, the closer it will be to the distribution of held out relevant items for that user. As seen in Table~\ref{tab:fit}, the results validate the idea that unimodal user embeddings do not model user interests as well as mixtures over multiple embeddings. Multiple embeddings yield a significant EMD improvement over a single embedding vector when evaluated on held-out items. Smoothing further decreases the EMD which we posit is due to the smoothed embedding mixtures being lower-variance estimates as they leverage engagement data from similar users in constructing the representations. These results suggest that the higher recall of smoothed mixtures is due to better user preferences modeling.

\begin{table}
    \vspace{-0.8cm}
    \centering
        \caption{Goodness of fit between user and held-out items as measured by earth mover's distance over a Euclidean embedding space. Lower EMD is better.}
        \label{tab:fit}
        \begin{tabular}{l c c c}
        \toprule
        \textbf{Approach} &HEP-TH & DBLP & Twitter-Follow\\
        \midrule
        Unimodal & 0.897 & 0.889  & 1.018  \\
        Mixture & 0.838 & 0.830  & 0.952  \\
        \knnembed & \textbf{0.811} & \textbf{0.808}  & \textbf{0.940}  \\
        \bottomrule
        \end{tabular}
\end{table}

\begin{figure*}
     \centering
     \begin{subfigure}[b]{0.315\textwidth}
         \centering
         \includegraphics[width=\textwidth]{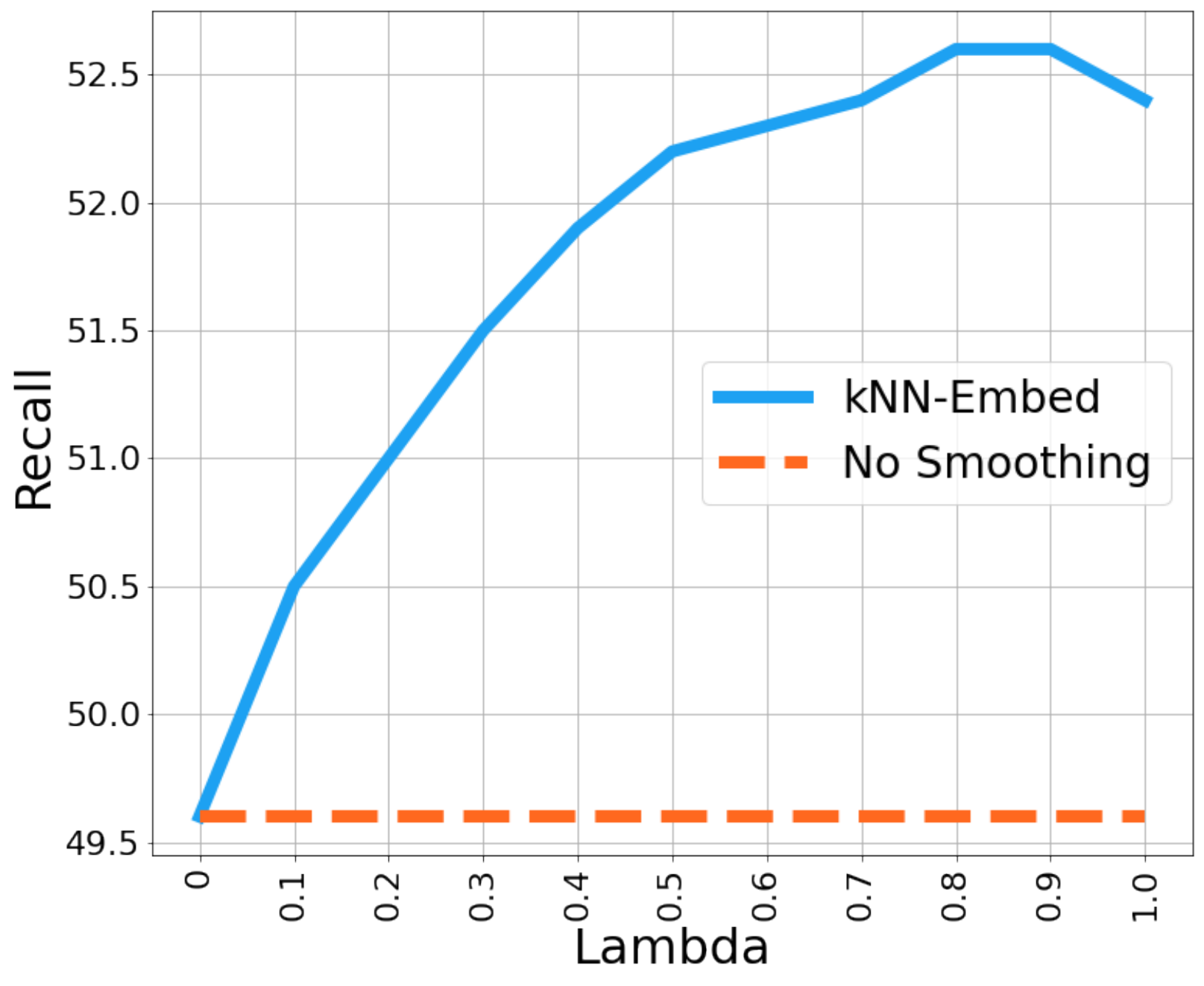}
         \caption{Varying lambda -- R@50.}
         \label{fig:lambda}
     \end{subfigure}
     \hfill
     \begin{subfigure}[b]{0.315\textwidth}
         \centering
         \includegraphics[width=\textwidth]{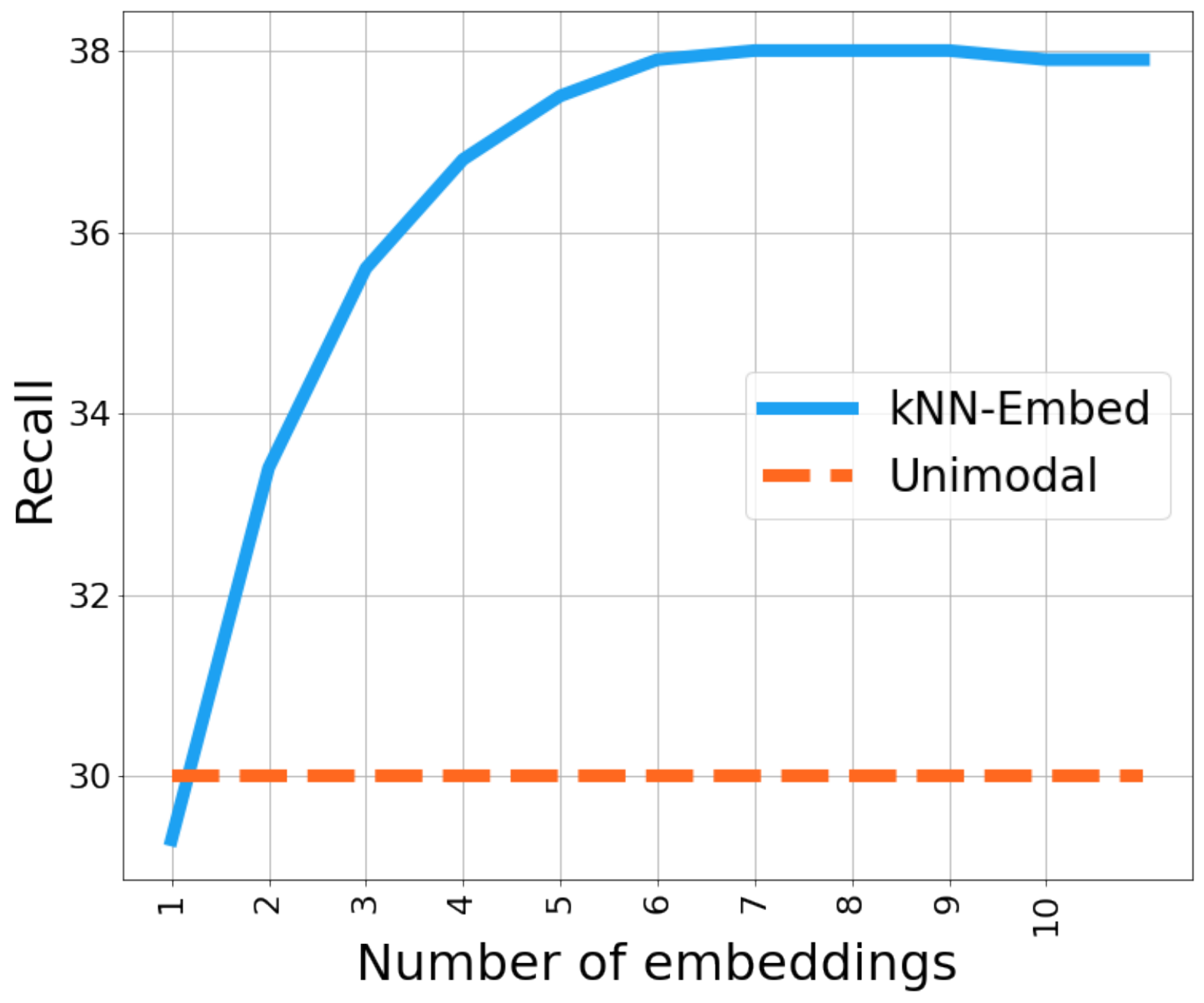}
         \caption{Varying mixtures R@20.}
         \label{fig:num_mixtures}
     \end{subfigure}
     \hfill
     \begin{subfigure}[b]{0.302\textwidth}
         \centering
         \includegraphics[width=\textwidth]{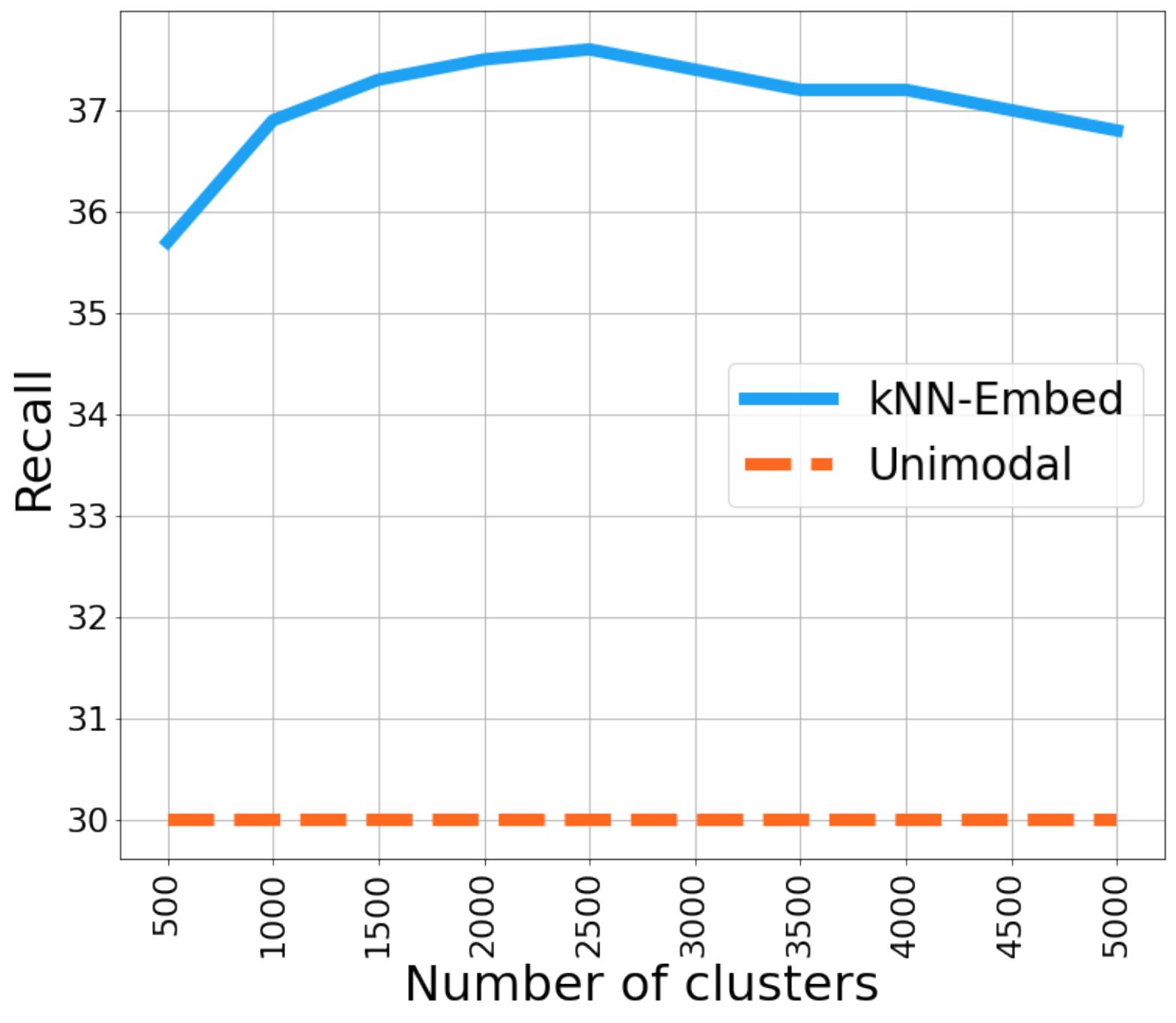}
         \caption{Varying clusters R@20.}
         \label{fig:num_clusters}
     \end{subfigure}
        \caption{We analyze the effect of three important hyper-parameters: (1) the $\lambda$ smoothing  (2) the number of embeddings in the mixture (3) the number of clusters for candidate retrieval in the HEP-TH dataset.}
        \label{fig:b}
    \vspace{-0.8cm}
\end{figure*}

\paragraph{\textbf{Hyper-parameter Sensitivity Analysis}:} We focus on recall as the \textit{sine qua non} of candidate retrieval problems and analyze hyper-parameters on HEP-TH. In Figure \ref{fig:lambda}, we vary the smoothing parameter $\lambda$ (same parameter for both the mixture probabilities and the cluster centroids) and see heavy smoothing improves performance significantly. This likely stems from the sparsity of HEP-TH where most papers have only a few citations. In Figure \ref{fig:num_mixtures}, we vary the number of embeddings (i.e., the mixture size) and notice improved performance saturating at six mixture components. Out of all the hyperparameters, this seems to be the critical one in achieving high recall. In practice, latency constraints can be considered when selecting the number of embeddings per user, explicitly making the trade-off between diversity and latency. Finally, in Figure \ref{fig:num_clusters}, we vary the number of k-means clusters; recall peaks at $k=2500$ and then decreases. HEP-TH is a small dataset with only 34,546 items; it is likely that generating a very large number of clusters leads to excessively fine-grained and noisy sub-divisions of the items.%

\vspace{-0.1cm}
\section{Conclusions}\label{sec:conclusions}
We present \knnembed, a method of transforming single user dense embeddings, into mixtures of embeddings, with the goal of better modeling user interests, increasing retrieval recall and diversity. This multi-embedding scheme represents a source entity with multiple distinct topical affinities by globally clustering items and aggregating the source entity’s engagements with clusters. Recognizing that user-item engagements may often be sparse, we propose a nearest-neighbor smoothing to enrich these mixture representation. Our smoothed mixture representation better models user preferences retrieving a diverse set of candidate items reflective of a user's multiple interests. This significantly improves recall on candidate retrieval tasks on three datasets including Twitter-Follow, a dataset we curate and release to the community.

\bibliographystyle{splncs04}
\bibliography{main}

\end{document}